\documentclass[conference, a4paper]{IEEEtran}

\usepackage{geometry}
\usepackage{graphicx}
\usepackage{color}
\usepackage{placeins}
\usepackage{float}
\usepackage{tabularx,colortbl}
\usepackage{amssymb}
\usepackage{amsmath}
\usepackage{units}
\usepackage{lipsum}
\usepackage{cuted} \stripsep -3pt plus 20pt minus 2pt
\DeclareMathOperator{\sinc}{sinc}
\usepackage{amsthm}
\usepackage{cite}
\usepackage{amsmath}
\usepackage{caption2}

\usepackage{cases,subeqnarray}
\usepackage{bm,multirow,bigstrut}
\usepackage{textcomp}
\usepackage{latexsym,bm}
\usepackage{booktabs}
\usepackage{xcolor}
\usepackage{mathtools}
\usepackage{dsfont}
\usepackage{extarrows}
\usepackage{subfigure}
\allowdisplaybreaks[4]
\usepackage{epsfig}
\usepackage{epstopdf} 
\usepackage{algorithm}
\usepackage{algpseudocode}

\theoremstyle{plain}

\newtheorem{rem}{Remark}

\usepackage{times} 
\IEEEoverridecommandlockouts

\begin{document}


\title{Performance Analysis of RIS-aided MISO Systems with EMI and Channel Aging}
\vspace{-20mm}
\author{
    \IEEEauthorblockN{Taoyu Song, Enyu Shi, Yu Lu, Yiyang Zhu, Jiayi Zhang, and Bo Ai,~\IEEEmembership{Fellow,~IEEE}}
    \IEEEauthorblockA{School of Electronic and Information Engineering, Beijing Jiaotong University, Beijing 100044, China}
    \thanks{This work was supported by the Fundamental Research Funds for the Central Universities under Grants 2023YJS015.}
    }


\vspace{-20mm}

\maketitle

\begin{abstract}
In this paper, we investigate a reconfigurable intelligent surface (RIS)-aided multiple-input single-output (MISO) system in the presence of electromagnetic interference (EMI) and channel aging with a Rician fading channel model between the base station (BS) and user equipment (UE). Specifically, we derive the closed-form expression for downlink spectral efficiency (SE) with maximum ratio transmission (MRT) precoding. The Monte-Carlo simulation supports the theoretical results, demonstrating that amplifying the weight of the line-of-sight (LoS) component in Rician fading channels can boost SE, while EMI has a detrimental impact. Furthermore, continuously increasing the number of RIS elements is not an optimal choice when EMI exists. Nonetheless, RIS can be deployed to compensate for SE degradation caused by channel aging effects. Finally, enlarging the RIS elements size can significantly improve system performance.
\end{abstract}

\IEEEpeerreviewmaketitle

\section{Introduction}
With the evolution of wireless communication technology, it is anticipated that 5G may not meet future demands for ultra-high data rates, ultra-low latency, and extensive coverage \cite{zhang2020cfmMIMO}. Therefore, in recent years, the global industry has started preliminary research on 6G and brought about significant advancements in areas such as millimeter-wave (mmWave) communications \cite{Ra2017MillimeterWave}, cell-free (CF) massive multiple-input multiple-output (MIMO) \cite{shi2022wireless}, reconfigurable intelligent surface (RIS) \cite{shi2023ris}, and ultra-dense network (UDN) \cite{An2016UDN}. RIS is a rising technology that can shape radio waves without using complex digital signal processing methods or active power amplifiers \cite{yan2022RIS}. Thus, it has recently been considered a new way to create reconfigurable wireless channels in some scenarios with hostile communication environments \cite{wu2019intelligent,zhang2021ris}. Due to the different fabrication, RIS exhibits distinct properties compared to traditional uniform linear arrays (ULAs). 

However, \cite{bj2022EMImode} demonstrates that natural electromagnetic interference (EMI) can significantly affect the system performance, highlighting the importance of taking EMI into account in RIS-aided systems. In \cite{Shi2023Uplink}, the authors consider the effect of EMI on the uplink spectral efficiency (SE) of user equipment (UE) in CF massive MIMO systems and it indicates that EMI significantly affects SE, and the influence of EMI becomes more pronounced when the components of RIS are sufficiently large. Meanwhile, \cite{Lu2023performence} highlights that the dynamic channel hasn't been thoroughly explored in current research, and most setups rely on the block fading model with roughly uniform channel features within coherent blocks. The decline in performance stems from the disparity between the real channel and the estimated channel due to user position shifts, thus necessitating a reevaluation of the influence of channel aging in RIS-aided systems.

Building upon prior research, this paper explores closed-form expressions for the downlink SE of RIS-aided multiple-input single-output (MISO) system with EMI over spatially correlated channels. We assume a Rician fading channel between the RIS and the base station (BS) while the channels from UE to RIS and UE to BS are Rayleigh fading. We employ a minimum mean square error (MMSE) estimator for channel state information (CSI) from the pilot signal, and maximum ratio transmission (MRT) precoding is used at BS. Then, we analyze the impact of various parameters on SE and validate the equation through Monte Carlo simulation. The results show that increasing line-of-sight (LoS) weight in the Rician fading channel improves SE, RIS can alleviate the effects of channel aging, and enlarging the RIS size can enhance system performance. However, EMI will have a significant negative impact when there are more RIS elements. Moreover, increasing transmission time and user movement speed will amplify the effects of channel aging.



\section{System Model}\label{se:model}
We assume that the BS has \(N_t\) antennas. The RIS consists of a controller and \(M\) elements capable of applying phase shifts to incoming signals. For the UE side, we consider a constant speed denoted as \(v\) for user movement while the BS and the RIS remain static. We assume that each coherence time block is represented as \(\tau _c\ \). In the uplink, we allocate  \(\tau _u\) symbols for channel estimation, while \(\tau _{\text{d}}=\tau _{\text{c}}-\tau _{\text{u}}\) symbols are for data transmission in the downlink. The channels are assumed to remain unchanged over symbol duration but vary across symbol time gaps. We assume independent Rayleigh fading models between the BS and UE, as well as between the RIS and UE, and Rician fading model is used between RIS and BS.

\subsection{Channel Model}
At the time of transmitting the $n$-th symbol, the channel matrix of the direct link between the BS and the UE can be expressed as \(\boldsymbol{G}_d\left[ n \right] =\left[ \boldsymbol{g}_{\text{d},1}\left[ n \right] ,...,\boldsymbol{g}_{\text{d},\text{k}}\left[ n \right],...,\boldsymbol{g}_{\text{d},\text{K}}\left[ n \right] \right] \in \mathbb{C}^{N_t\times K}\) where \(\boldsymbol{g}_{d,k}\left[ n \right] \in \mathbb{C}^{N_t}\) denotes the channel between UE \(k\) and the BS. Specially, \(\boldsymbol{g}_{d,k}\left[ n \right] \) can be expressed as \(\boldsymbol{g}_{d, k}\left[ n \right] =\sqrt{\beta _{d, k}}\boldsymbol{h}_{d, k}\left[ n \right] , \forall k\in \left\{ 1, . . . , K \right\} \) where \(\boldsymbol{h}_{d, k}\left[ n \right] \in \mathbb{C}^{N_t}\) denotes the small-scale fading coefficient between UE $k$ and the BS, \(\beta _{d, k}\) denotes the path loss of the direct path. For the channel between the RIS and the BS, we can express as \cite{zhang2022reconfigurable}
{\setlength\abovedisplayskip{0.1cm}
\setlength\belowdisplayskip{0.1cm}
\begin{equation}
\boldsymbol{G}_{br}=\sqrt{\beta _{br}}\left( \sqrt{\frac{\kappa}{\kappa +1}}\boldsymbol{\bar{g}}_{br}+\sqrt{\frac{1}{\kappa +1}}\boldsymbol{\tilde{g}}_{br} \right) \in \mathbb{C}^{N_t\times M},
\end{equation}}%
where \(\boldsymbol{\bar{g}}_{br}=\left[ \boldsymbol{\bar{g}}_{br1}, . . . , \boldsymbol{\bar{g}}_{brm}, . . . , \boldsymbol{\bar{g}}_{brM} \right] \in \mathbb{C}^{N_t\times M}\) denotes the deterministic LoS component, while  \(\boldsymbol{\bar{g}}_{brm}\in \mathbb{C}^{N_t}\) can be expressed as \(\boldsymbol{\bar{g}}_{brm}=\left[ \alpha \exp \left( j\theta _1 \right) , . . . , \alpha \exp \left( j\theta _n \right) , . . . , \alpha \exp \left( j\theta _{N_t} \right) \right] ^T\) where \(\alpha \in \left[ 0, 1 \right] \) and \(\theta _n\in \left[ 0, 2\pi \right] \) represents the amplitude and phase of the signal. We assume that the amplitude of the signal reaches its maximum. Meanwhile, \(\boldsymbol{\tilde{g}}_{br}\) is the non-line of sight (NLoS) component. \(\beta _{br}\) denotes the corresponding path loss. At the time of transmitting the \(n\)-th symbol, the channel matrix of the direct link between the UE and the RIS can be expressed as \(\boldsymbol{g}_{r, k}\left[ n \right] =\left[ g_{r, k1}\left[ n \right] , . . . ,  g_{r, kM}\left[ n \right] \right] ^T\in \mathbb{C}^{M}\) where \(g_{r, km}\left[ n \right] =\sqrt{\beta _{r, k}}h_{r, km}\left[ n \right] \) denotes the channel between UE \(k\) and \(m\)-th element of RIS. We denote the channel between the RIS and the UE as \(\boldsymbol{G}_{r, ue}\left[ n \right] =\left[ \boldsymbol{g}_{r, 1}\left[ n \right] , . . . , \boldsymbol{g}_{r, k}\left[ n \right] \right] \in \mathbb{C}^{M\times K}\). And \(\boldsymbol{g}_{r, k}\sim \mathcal{N}\left( \mathbf{0}, A\beta _{r, k}\boldsymbol{R}\right) \), where \(\beta _{r, k}\) denotes the corresponding path loss. \(d_H\) and \(d_V\) represent the horizontal width and the vertical height of a RIS element, and \(A=d_Hd_V\) denotes the area of the RIS element. \(\boldsymbol{R}\in \mathbb{C}^{M\times M}\) denotes the correlation matrix which has the \(\left( n, m\right) \)-th element as \(\left[ \boldsymbol{R} \right] _{n, m}=\sinc\left( 2\lVert u_n-u_m \rVert/{\lambda} \right) \) where \(\sinc\left( x \right) =\sin \left( \pi x \right) /\left( \pi x \right) \) and \(\lambda\) denotes the carrier wavelength. The position of the \(m\)-th element with respect to the origin is \(\boldsymbol{u}_m=\left( 0, mod\left( m-1, M_H \right) d_H, \lfloor \left( m-1 \right) /M_H \rfloor d_V \right) ^T, m\in \left[ 0, M \right]\) , where  \(M_H\) and \(M_V\) are the
numbers of elements at RIS in each row and column, such that \(M=M_H\times M_V\) \cite{bj2021Rayleigh_RIS_model}. Therefore, the cascade channel can be represented as \(\boldsymbol{G}_{c, k}\left[ n \right] =\boldsymbol{G}_{br}\boldsymbol{\varPhi }\left[ n \right] \boldsymbol{g}_{r, k}\left[ n \right] \) , where \(\boldsymbol{\varPhi }\left[ n \right] =\text{diag}\left\{ \phi \exp \left( j\theta _1 \right) , . . . , \phi \exp \left( j\theta _M \right) \right\} \in \mathbb{C}^{M\times M}\) denotes the RIS reflection matrix, \(\phi \in \left[ 0, 1 \right] \) and \(\theta _m\in \left[ 0, 2\pi \right] \) represents the amplitude and phase of the reflected signal. Due to the sparsity and serious path loss of the mmWave channel, we only consider the signal reflected by the RIS for the first time \cite{Lu2023performence}. Same as above, we assume that \(\alpha =1\). We can also express \(\boldsymbol{G}_{c, k}\left[ n \right]\) as \(\boldsymbol{G}_{c, k}\left[ n \right] =\boldsymbol{g}_{ck}\left[ n \right] \boldsymbol{v}\left[ n \right] \), where \(\boldsymbol{G}_{ck}\left[ n \right] =\left[ \boldsymbol{g}_{ck1}\left[ n \right] , . . . , \boldsymbol{g}_{ckM}\left[ n \right] \right] \in \mathbb{C}^{N_t\times M}\) represents the cascade channel matrix and \(\boldsymbol{v}\left[ n \right] =\left[ \phi \exp \left( j\theta _1 \right) , . . . , \phi \exp \left( j\theta _M \right) \right] \in \mathbb{C}^{M\times 1}\) denotes the RIS reflection matrix . Moreover, \(\boldsymbol{g}_{ckm}\left[ n \right] \) can be expressed as \(\boldsymbol{g}_{ckm}\left[ n \right] =\boldsymbol{g}_{brm}g_{r, km}\left[ n \right]\).

\subsection{Electromagnetic Interference Model}
The EMI is an uncontrolled electromagnetic wave generated by electronic devices, which can cause interference to other devices. In \cite{bj2022EMImode}, the author demonstrates that the EMI at the RIS significantly affects communication quality. Especially, the EMI is donated as \(\boldsymbol{n}\sim \mathcal{N}\left( \mathbf{0}, A\sigma_e ^2\boldsymbol{R} \right) \) where \(\sigma_e ^2\) is the EMI power at RIS. To further investigate the effects of EMI, we define \(\rho =P_{\tau _p}\beta _{br}/\sigma_{e}^2\) which corresponds to the ratio between the received signal power and EMI power at each element of the RIS \cite{bj2022EMImode}.
\vspace{-2mm}
\begin{rem}
For the sinc function, the output is zero only when the argument is a non-zero integer, implying that the magnitude of  \(\lVert u_n-u_m \rVert\) must be an integer multiple of \(\lambda/2\). This means that the RIS elements should be aligned along a straight line with intervals that are integer multiples of \(\lambda/2\), which is impossible to achieve with a two-dimensional RIS in practice. Thus, it is essential to take into account the impact of EMI with spatial correlation at the RIS elements.
\end{rem}
\vspace{-3mm}
\subsection{ Channel Estimation}
We use \(\tau_p\) pilot sequences to estimate channel, in every coherence time block. The pilot signal emitted by UE \(k\), which is represented by \(\boldsymbol{\varphi }_k\in \mathbb{C}^{\tau _p}\), meets the condition that \(\lVert \boldsymbol{\varphi }_k \rVert ^2=1\). Thus, the pilot matrix transmitted by the \(K\) UE is mutually orthogonal as \(\boldsymbol{\varPsi }=\left[ \boldsymbol{\varphi }_1, . . . , \boldsymbol{\varphi }_k \right] \in \mathbb{C}^{\tau _p\times K}\), which satisfies \(\boldsymbol{\varPsi }^{\boldsymbol{H}}\boldsymbol{\varPsi }=\boldsymbol{I}_{\boldsymbol{K}}\). The pilot signals from different UE satisfy \(\boldsymbol{\varphi }_{i}^{H}\boldsymbol{\varphi }_j=0, \forall i\ne j\). Considering the EMI, the received signal \(\boldsymbol{G}_{r, k}\left[ n \right]\) at the RIS is \(\boldsymbol{G}_{r, k}\left[ n \right] =\boldsymbol{g}_{r, k}\left[ n \right] \boldsymbol{\varphi}_{k}^{T}\left[ n \right] +\boldsymbol{N}\) where \(\boldsymbol{N}\in \mathbb{C}^{M\times \tau _p}\) represents the additive EMI at RIS. We perform linear MMSE estimates for direct and cascade links. Initially, we obtain channel vector estimates, then we derive channel characteristics at subsequent time points based on temporal correlation. During pilot signal transmission, we assume constant direct and cascade channels, allowing us to disregard the effects of channel aging \cite{Lu2023performence}. For the channels between UE and BS, the signal received by the BS can be presented as
{
\setlength\abovedisplayskip{0.1cm}
\setlength\belowdisplayskip{0.1cm}
\begin{align}
\boldsymbol{Y}_{d}^{p}\left[ 0 \right] =\sqrt{P_{\tau _p}}\boldsymbol{G}_d\left[ 0 \right] \boldsymbol{\varPsi }^H+\boldsymbol{Z}_{d}^{p}\left[ 0 \right],
\label{eq8}
\end{align}
}%
where \(P_{\tau _p}=\tau _pP_{\tau _u}\) represents the pilot signal power, \(P_{\tau _u}\) is the average uplink transmission power per UE. \(\boldsymbol{Z}_{d}^{p}\left[ 0 \right] \sim \mathcal{N}\left( \mathbf{0}, \sigma _{d}^{2}\boldsymbol{I}_{N_t} \right) \in \mathbb{C}^{N_t\times \tau _p}\) represents the noise received at the BS at the initial state, where \(\sigma _{d}^{2}\) represents the variance of the noise. Then, we multiply \eqref{eq8} by \(1/{\sqrt{P_{\tau _p}}}\boldsymbol{\varphi }_k\) and we get the direct link as
{
\setlength\abovedisplayskip{0.1cm}
\setlength\belowdisplayskip{0.1cm}
\begin{align}
\boldsymbol{\tilde{y}}_{d, k}^{p}\left[ 0 \right] =\boldsymbol{g}_{d, k}\left[ 0 \right] +\frac{1}{\sqrt{P_{\tau _p}}}\boldsymbol{Z}_{d}^{p}\left[ 0 \right] \boldsymbol{\varphi }_k.
\end{align}
}%
The optimal \(\boldsymbol{g}_{d, k}\left[ 0 \right]\) can be obtained using the MMSE estimation method, which minimizes the mean squared error \(\mathbb{E}\left( \lVert \boldsymbol{g}_{d, k}\left[ 0 \right] -\boldsymbol{\hat{g}}_{d, k}\left[ 0 \right] \rVert ^2 \right) \) \cite{kun2021estimation}. So the estimated \(\boldsymbol{g}_{d, k}\left[ 0 \right]\) can be expressed as \cite{Lu2023performence}
{
\setlength\abovedisplayskip{0.1cm}
\setlength\belowdisplayskip{0.1cm}
\begin{equation}
\boldsymbol{\hat{g}}_{d, k}\left[ 0 \right] =\left( 1+\frac{\sigma _{d}^{2}}{P_{\tau _p}\beta _{d, k}} \right) ^{-1}\boldsymbol{\tilde{y}}_{d, k}^{p}\left[ 0 \right] .
\end{equation}
}%
Thus, \(\boldsymbol{g}_{d, k}\left[ 0 \right]\) can be expressed as \(\boldsymbol{g}_{d, k}\left[ 0 \right] =\boldsymbol{\hat{g}}_{d, k}\left[ 0 \right] +\boldsymbol{\tilde{g}}_{d, k}\left[ 0 \right] \) where \(\boldsymbol{\hat{g}}_{d, k}\left[ 0 \right] \sim \mathcal{N}\left( \mathbf{0}, \frac{P_{\tau _p}\beta _{d, k}^{2}}{\sigma _{d}^{2}+P_{\tau _p}\beta _{d, k}}\boldsymbol{I}_{N_t} \right) \). And \(\boldsymbol{\tilde{g}}_{d, k}\left[ 0 \right] \sim \mathcal{N}\left( \mathbf{0}, \frac{\sigma _{d}^{2}\beta _{d, k}}{\sigma _{d}^{2}+P_{\tau _p}\beta _{d, k}}\boldsymbol{I}_{N_t} \right) \) is the estimation error which is uncorrelated with \(\boldsymbol{\hat{g}}_{d, k}\left[ 0 \right]\). We can get the cascade link between the BS and UE without the reflection matrix as
{\setlength\abovedisplayskip{0.1cm}
\setlength\belowdisplayskip{0.1cm}
\begin{align}
\boldsymbol{Y}_{c}^{p}\left[ 0 \right] =\sqrt{P_{\tau _p}}\left( \boldsymbol{G}_d\left[ 0 \right] \boldsymbol{\varPsi }^H+\boldsymbol{G}_{br} \boldsymbol{G}_r\left[ 0 \right] \right) +\boldsymbol{Z}_{c}^{p}\left[ 0 \right] ,\label{eq13}
\end{align}
}%
where \(\boldsymbol{G}_r\left[ 0 \right]\) can be expressed as \(\boldsymbol{G}_r\left[ 0 \right] =\boldsymbol{G}_{r, ue}\left[ 0 \right] \boldsymbol{\varPsi }^H+\boldsymbol{N}\). Then, we multiply \eqref{eq13} by\(1/{\sqrt{P_{\tau _p}}}\boldsymbol{\varphi }_k\) and we get the cascade link  as
{\setlength\abovedisplayskip{0.1cm}
\setlength\belowdisplayskip{0.1cm}
\begin{align}
\boldsymbol{\tilde{y}}_{c, k}^{p}\!\left[ 0 \right]\!\!=\!\boldsymbol{g}_{d, k}\!\left[ 0 \right] \!+\!\boldsymbol{g}_{ck}\!\left[ 0 \right] \!+\!\boldsymbol{G}_{br}\boldsymbol{N\varphi }_k+\frac{1}{\sqrt{P_{\tau _p}}}\boldsymbol{Z}_{c}^{p}\!\left[ 0 \right] \boldsymbol{\varphi }_k.
\end{align}
}%
Here, \(\boldsymbol{Z}_{c}^{p}\left[ 0 \right] \sim \mathcal{N}\left( \mathbf{0}, \sigma _{c}^{2}\boldsymbol{I}_{N_t} \right) \in \mathbb{C}^{N_t\times \tau _p}\) represents the noise received at the BS at the initial state, where \(\sigma _{c}^{2}\) represents the variance of the noise. After ignoring the channel estimation error, we obtain the cascade link as
{\setlength\abovedisplayskip{0.1cm}
\setlength\belowdisplayskip{0.1cm}\begin{equation}
\boldsymbol{\tilde{y}}_{c, k}^{p}\left[ 0 \right] =\boldsymbol{\tilde{g}}_{d, k}\left[ 0 \right] +\boldsymbol{g}_{ck}\left[ 0 \right] +\boldsymbol{G}_{br}\boldsymbol{N\varphi }_k+\frac{1}{\sqrt{P_{\tau _p}}}\boldsymbol{Z}_{c}^{p}\left[ 0 \right] \boldsymbol{\varphi }_k.
\end{equation}
}%
So the MMSE estimation of \(\boldsymbol{g}_{ck}\left[ 0 \right]\) can be expressed as
{
\setlength\abovedisplayskip{0.1cm}
\setlength\belowdisplayskip{0.1cm}
\begin{equation}
\boldsymbol{\hat{g}}_{ck}\left[ 0 \right] =\xi _{c, k}\left( \frac{\sigma _{d}^{2}\beta _{d, k}}{\sigma _{d}^{2}+P_{\tau _p}\beta _{d, k}}+\xi _{c, k}+Q+\frac{\sigma _{c}^{2}}{P_{\tau _p}} \right) ^{-1}\boldsymbol{\tilde{y}}_{c, k}^{p}\left[ 0 \right] ,
\end{equation}
}%
where \(\xi _{c, k}=A\beta _{r, k}\beta _{br}\left( \kappa /\left( \kappa +1 \right) +A/\left( \kappa +1 \right) \right) \) and \(Q=A\sigma_e ^2tr\left( \boldsymbol{R}_e \right) \beta _{br}\left( \kappa /\left( \kappa +1 \right) +A/\left( \kappa +1 \right) \right)\). Similarly, \(\boldsymbol{g}_{ck}\left[ 0 \right]\) can be expressed as \(\boldsymbol{g}_{ck}\left[ 0 \right] =\boldsymbol{\hat{g}}_{ck}\left[ 0 \right] +\boldsymbol{\tilde{g}}_{ck}\left[ 0 \right] \) where \(\boldsymbol{\hat{g}}_{ck}\left[ 0 \right] \sim \mathcal{N}\left( \mathbf{0}, \frac{\xi _{c, k}\sigma _{e1, k}^{2}}{\sigma _{e1, k}^{2}+\sigma _{e2, k}^{2}+\sigma _{e3, k}^{2}+\sigma _{d}^{2}\sigma _{c}^{2}}\boldsymbol{I}_{N_t} \right)\). And \(\boldsymbol{\tilde{g}}_{ck}\left[ 0 \right] \sim \mathcal{N}\left( \mathbf{0}, \frac{\xi _{c, k}\left( \sigma _{e2, k}^{2}+\sigma _{e3, k}^{2} \right)}{\sigma _{e1, k}^{2}+\sigma _{e2, k}^{2}+\sigma _{e3, k}^{2}+\sigma _{d}^{2}\sigma _{c}^{2}}\boldsymbol{I}_{N_t} \right) \) is the estimation error which is uncorrelated with \(\boldsymbol{\hat{g}}_{ck}\left[ 0 \right]\), where \(\sigma _{e1, k}^{2}\triangleq P_{\tau _p}\xi _{c, k}\left( \sigma _{d}^{2}+P_{\tau _p}\beta _{d, k} \right) , \sigma _{e2, k}^{2}\triangleq P_{\tau _p}\beta _{d, k}\left( \sigma _{d}^{2}+\sigma _{c}^{2} \right) , \sigma _{e3, k}^{2}\triangleq Q\left( \sigma _{d}^{2}+P_{\tau _p}\beta _{d, k} \right)\).

\subsection{Channel Aging}
Due to the relative position change between the UE and the BS causing channel aging, the CSI is usually not perfectly estimated. Therefore, we must consider the impact of outdated CSI on system performance. According to \cite{Lu2023performence}, \(\boldsymbol{g}_{d, k}\left[ n \right]\) and \(\boldsymbol{g}_{r, k}\left[ n \right]\) can be represented by their initial state, respectively
 {
\setlength\abovedisplayskip{0.1cm}
\setlength\belowdisplayskip{0.1cm}
\begin{align}
&\boldsymbol{g}_{d, k}\left[ n \right] =\rho _0\left[ n \right] \boldsymbol{g}_{d, k}\left[ 0 \right] +\bar{\rho}_0\left[ n \right] \boldsymbol{e}_{d, k}\left[ n \right] ,\label{eq20}\\
&g_{r, km}\left[ n \right] =\rho _1\left[ n \right] g_{r, km}\left[ 0 \right] +\bar{\rho}_1\left[ n \right] e_{r, km}\left[ n \right] .\label{eq21}
\end{align}
 }%
According to the Jakes' model \cite{Lu2023performence}, the temporal correlation coefficient is donated as \(\rho _i\left[ n \right] =J_0\left( 2\pi nf_DT_s \right) \left( i\in \left\{ 0, 1 \right\} \right)\) \cite{Anastssions2017Impact}, where \(J_0\left( \cdot \right)\) is the initial class of zeroth-order Bessel function, \(f_c\) is the carrier frequency, \(c\) is the velocity of light, so \(f_D=f_cv/c\) is the maximum Doppler shift and \(T_s\) is the sample interval. We assume \(\rho _i\left[ n \right]\) is known at the BS. Moreover, \(\bar{\rho}_i\left[ n \right] =\sqrt{1-\rho _{i}^{2}\left[ n \right]}\) donates the coefficient of the current channel part independent of the initial state of the channel. This part satisfies that \(\boldsymbol{e}_{d, k}\left[ n \right] \sim \mathcal{N}\left( \mathbf{0}, \beta _{d, k}\boldsymbol{I}_{N_t} \right) \), and \(e_{r, km}\left[ n \right] \sim \mathcal{N}\left( 0, 1 \right)\), respectively. 
Thus, \(\boldsymbol{g}_{ckm}\left[ 0 \right] \) can be written as \( \boldsymbol{g}_{ckm}\left[ 0 \right] =\rho _1\left[ n \right] \boldsymbol{g}_{brm}g_{r, km}\left[ 0 \right] +\bar{\rho}_1\left[ n \right] \boldsymbol{g}_{brm}e_{r, km}\left[ n \right] \). We define that \(\boldsymbol{e}_{ckm}\left[ n \right] \triangleq \boldsymbol{g}_{brm}e_{r, km}\left[ n \right] \), so \(\boldsymbol{e}_{ck}\left[ n \right] \sim \mathcal{N}\left( \mathbf{0}, \xi _{c, k}\boldsymbol{I}_{N_t} \right)\). 

After considering the combined effect of estimation error and channel aging we can obtain the expressions for cascading links as
{ \setlength\abovedisplayskip{0.1cm}
\setlength\belowdisplayskip{0.1cm}\begin{align}
\boldsymbol{g}_{ck}\left[ 0 \right] =\rho _1\left[ n \right] \left( \boldsymbol{\hat{g}}_{ck}\left[ 0 \right] +\boldsymbol{\tilde{g}}_{ck}\left[ 0 \right] \right) +\bar{\rho}_1\left[ n \right] \boldsymbol{e}_{ck}\left[ n \right] ,
 \label{eq23}
\end{align}
 }%
where \(\boldsymbol{\tilde{e}}_{ck}\left[ n \right] \triangleq \rho _1\left[ n \right] \boldsymbol{\tilde{g}}_{ck}\left[ 0 \right] +\bar{\rho}_1\left[ n \right] \boldsymbol{e}_{ck}\left[ n \right]\), and \(\boldsymbol{\tilde{e}}_{ck}\left[ n \right] \sim \mathcal{N}\left( \mathbf{0}, \left( \xi _{c, k}-\frac{\rho _{1}^{2}\left[ n \right] \xi _{c, k}\sigma _{e1, k}^{2}}{\sigma _{e1, k}^{2}+\sigma _{e2, k}^{2}+\sigma _{e3, k}^{2}+\sigma _{d}^{2}\sigma _{c}^{2}} \right) \boldsymbol{I}_{N_t} \right)\). Similarly, the direct link can be expressed as
 { \setlength\abovedisplayskip{0.1cm}
\setlength\belowdisplayskip{0.1cm}\begin{align}
\boldsymbol{g}_{d, k}\left[ 0 \right]=\rho _0\left[ n \right] \boldsymbol{\hat{g}}_{d, k}\left[ 0 \right] +\boldsymbol{\tilde{e}}_{d, k}\left[ n \right], \label{eq24}
\end{align}
 }%
 where \(\boldsymbol{\tilde{e}}_{d, k}\left[ n \right] \triangleq \rho _0\left[ n \right] \boldsymbol{\tilde{g}}_{d, k}\left[ 0 \right] +\bar{\rho}_0\left[ n \right] \boldsymbol{e}_{d, k}\left[ n \right]\), and \(\boldsymbol{\tilde{e}}_{d, k}\left[ n \right] \sim \mathcal{N}\left( \mathbf{0}, \left( \beta _{d, k}-\frac{\rho _{0}^{2}\left[ n \right] P_{\tau _p}\beta _{d, k}^{2}}{\sigma _{d}^{2}+P_{\tau _p}\beta _{d, k}} \right) \boldsymbol{I}_{N_t} \right)\).

\section{Performance Analysis and Useful Insights}\label{se:performance}
Based on the channel reciprocity \cite{Tr2013reciprocity} and combined with the channel aging, we can express the received signal at UE \(k\) in the phase of downlink signal transmission as
{\setlength\abovedisplayskip{0.1cm}
\setlength\belowdisplayskip{0.1cm}\begin{align}
y_k\left[ n \right] &=\left( \boldsymbol{g}_{d, k}^{H}\left[ n \right] +\boldsymbol{v}^H\left[ n \right] \boldsymbol{g}_{ck}^{H}\left[ n \right] \right) \boldsymbol{x}\left[ n \right] \nonumber\\
&+ \boldsymbol{g}_{r, k}^{H}\left[ n \right] \boldsymbol{\varPhi }^H\left[ n \right] \boldsymbol{u}+\omega _k\left[ n \right]
,
\end{align}
}%
where \(\boldsymbol{x}\left[ n \right] =\boldsymbol{F}\left[ n \right] \boldsymbol{s}\left[ n \right]\)represents the signal transmitted by the BS with power \(P_T\geqslant tr\left( \boldsymbol{F}^H\left[ n \right] \boldsymbol{F}\left[ n \right] \right) =\mathbb{E}\left( \lVert \boldsymbol{x}\left[ n \right] \rVert ^2 \right)\), and \(\boldsymbol{u}\sim \mathcal{N}\left( \mathbf{0}, A\sigma ^2\boldsymbol{R} \right) \in \mathbb{C}^M\) denotes the additive EMI noise. \(\boldsymbol{F}\left[ n \right] =\left[ \boldsymbol{f}_1\left[ n \right] , . . . , \boldsymbol{f}_K\left[ n \right] \right] \in \mathbb{C}^{N_t\times K}\) denotes the precoding matrix and \(\boldsymbol{s}\left[ n \right] =\left[ s_1\left[ n \right] , . . . , s_K\left[ n \right] \right] ^T\in \mathbb{C}^K\) donates received signal vector at UE, which satisfies \(\mathbb{E}\left( \boldsymbol{s}\left[ n \right] \right) =0\) and \(\mathbb{E}\left( \boldsymbol{s}\left[ n \right] \boldsymbol{s}^H\left[ n \right] \right) =\boldsymbol{I}_K\). \(\omega _k\left[ n \right] \) denotes additive white Gaussian noise at UE \(k\), which satisfies \(\omega _k\left[ n \right] \sim \mathcal{N}\left( 0, \sigma _{k}^{2} \right) \). We define \(\boldsymbol{G}_k\left[ n \right] \triangleq \boldsymbol{g}_{d, k}^{H}\left[ n \right] +\boldsymbol{v}^H\left[ n \right] \boldsymbol{g}_{ck}^{H}\left[ n \right]\), so we can get
{\setlength\abovedisplayskip{0.1cm}
\setlength\belowdisplayskip{0.1cm}\begin{align}
\boldsymbol{G}_k\left[ n \right] &=\rho _0\left[ n \right] \boldsymbol{\hat{g}}_{d, k}^{H}\left[ 0 \right] +\boldsymbol{\tilde{e}}_{d, k}^{H}\left[ n \right] \nonumber\\
&+\boldsymbol{v}^H\left[ n \right] \left( \rho _1\left[ n \right] \boldsymbol{\hat{g}}_{ck}^{H}\left[ 0 \right] +\boldsymbol{\tilde{e}}_{ck}^{H}\left[ n \right] \right) ,
\label{eq26}
\end{align}
}%
where we define \(\boldsymbol{\bar{G}}_k\left[ n \right] \triangleq \rho _0\left[ n \right] \boldsymbol{\hat{g}}_{d, k}^{H}\left[ 0 \right] +\rho _1\left[ n \right] \boldsymbol{v}^H\left[ n \right] \boldsymbol{\hat{g}}_{ck}^{H}\left[ 0 \right] \) and \(\boldsymbol{G}_{e, k}\left[ n \right] \triangleq \boldsymbol{\tilde{e}}_{d, k}^{H}\left[ n \right] +\boldsymbol{v}^H\left[ n \right] \boldsymbol{\tilde{e}}_{ck}^{H}\left[ n \right]\). Therefore, the received signal at UE \(k\) can be expressed as
 {\setlength\abovedisplayskip{0.1cm}
\setlength\belowdisplayskip{0.1cm}\begin{align}
y_k\left[ n \right] &=\mathbb{E}\left( \boldsymbol{G}_k\left[ n \right] \boldsymbol{f}_k\left[ n \right] \right) s_k\left[ n \right] +\sum_{j=1, j\ne k}^K{\boldsymbol{G}_k\left[ n \right] \boldsymbol{f}_j\left[ n \right] s_j\left[ n \right]}\nonumber\\
&+\boldsymbol{G}_k\left[ n \right] \boldsymbol{f}_k\left[ n \right] s_k\left[ n \right] -\mathbb{E}\left( \boldsymbol{G}_k\left[ n \right] \boldsymbol{f}_k\left[ n \right] \right) s_k\left[ n \right] \nonumber\\
&+\boldsymbol{g}_{r, k}^{H}\left[ n \right] \boldsymbol{\varPhi }^H\left[ n \right] \boldsymbol{u}+\omega _k\left[ n \right].
\end{align}
 }%
Using the MRT precoding algorithm, we assume that the CSI known at the BS at time \(n\) is \(\boldsymbol{G}_k\left[ n \right]\). The precoding matrix can be expressed as \cite{Lu2023performence} \(\boldsymbol{F}\left[ n \right] =\zeta \left[ n \right] \left( \boldsymbol{\bar{G}}\left[ n \right] \right) ^H \label{eq28}\), where \(\zeta ^2\left[ n \right] \triangleq 1/tr\left( \boldsymbol{\bar{G}}\left[ n \right] \left( \boldsymbol{\bar{G}}\left[ n \right] \right) ^H \right)\) denotes the normalized coefficient of the precoding matrix and \(\boldsymbol{\bar{G}}\left[ n \right] =\left[ \boldsymbol{\bar{G}}_1\left[ n \right] , . . . , \boldsymbol{\bar{G}}_K\left[ n \right] \right]^{T}\). According to the use-and-then-forget capacity bound \cite{Lu2023performence}, the minimum SE achievable by UE \(k\) is limited by \(R_k=\frac{1}{\tau _c}\sum_{n=1}^{\tau _c-\tau _u}{\log _2\left( 1+\gamma _k\left[ n \right] \right)},\)
where \(\gamma _k\left[ n \right]\) is the SINR of the UE \(k\) at time \(n\). \(\gamma _k\left[ n \right]\) can be expressed as \eqref{eq30} at the top of the next page.
{\setlength\abovedisplayskip{0cm}
\setlength\belowdisplayskip{0cm}\begin{figure*}
\begin{align}
\gamma _k\left[ n \right] =&\frac{\underset{I_{0. kn}}{\underbrace{\left| \mathbb{E}\left( \boldsymbol{G}_k\left[ n \right] \boldsymbol{f}_k\left[ n \right] \right) \right|^2}}}{\underset{I_{1, kn}}{\underbrace{Var\left( \boldsymbol{G}_k\left[ n \right] \boldsymbol{f}_k\left[ n \right] \right) }}+\underset{I_{2, kn}}{\underbrace{\sum_{j=1, j\ne k}^K{\mathbb{E}\left( \left| \boldsymbol{G}_k\left[ n \right] \boldsymbol{f}_j\left[ n \right] \right|^2 \right)}}}+\underset{I_{3, kn}}{\underbrace{\mathbb{E}\left( \left| \boldsymbol{g}_{r, k}^{H}\left[ n \right] \boldsymbol{\varPhi }^H\left[ n \right] \boldsymbol{u} \right|^2 \right) }}+\underset{I_{4. kn}}{\underbrace{\mathbb{E}\left( \left| \omega _k\left[ n \right] \right|^2 \right) }}}.\label{eq30}
\end{align}
\underline{\hspace{\textwidth}}
\end{figure*}}

\textbf{Corollary 1:}\\
{\setlength\abovedisplayskip{0.1cm}
\setlength\belowdisplayskip{0.1cm}\begin{align}
I_{0. kn}&=\zeta ^2\left[ n \right] \left( N_t\left( \frac{\rho _{0}^{2}\left[ n \right] P_{\tau _p}\beta _{d, k}^{2}}{\sigma _{d}^{2}+P_{\tau _p}\beta _{d, k}} \right. \right. 
\nonumber\\
&\left. \left. +\frac{M\rho _{1}^{2}\left[ n \right] \xi _{c, k}\sigma _{e1, k}^{2}}{\sigma _{e1, k}^{2}+\sigma _{e2, k}^{2}+\sigma _{e3, k}^{2}+\sigma _{d}^{2}\sigma _{c}^{2}} \right) \right) ^2,
\\
I_{1. kn}&=\zeta ^2\left[ n \right] N_t\left( \frac{M\rho _{1}^{2}\left[ n \right] \xi _{c, k}\sigma _{e1, k}^{2}}{\sigma _{e1, k}^{2}+\sigma _{e2, k}^{2}+\sigma _{e3, k}^{2}+\sigma _{d}^{2}\sigma _{c}^{2}} \right. 
\nonumber\\
&\left. +\frac{\rho _{0}^{2}\left[ n \right] P_{\tau _p}\beta _{d, k}^{2}}{\sigma _{d}^{2}+P_{\tau _p}\beta _{d, k}} \right) \times \left( \left( \beta _{d, k}-\frac{\rho _{0}^{2}\left[ n \right] P_{\tau _p}\beta _{d, k}^{2}}{\sigma _{d}^{2}+P_{\tau _p}\beta _{d, k}} \right) \right. \nonumber\\
&\left. +M\left( \xi _{c, k}-\frac{\rho _{1}^{2}\left[ n \right] \xi _{c, k}\sigma _{e1, k}^{2}}{\sigma _{e1, k}^{2}+\sigma _{e2, k}^{2}+\sigma _{e3, k}^{2}+\sigma _{d}^{2}\sigma _{c}^{2}} \right) \right) ,
\\
I_{2. kn}&=\sum_{j=1, j\ne k}^K{\zeta ^2\left[ n \right] N_t}\left( \frac{M\rho _{1}^{2}\left[ n \right] \xi _{c, k}\sigma _{e1, j}^{2}}{\sigma _{e1, j}^{2}+\sigma _{e2, j}^{2}+\sigma _{e3, j}^{2}+\sigma _{d}^{2}\sigma _{c}^{2}} \right. 
\nonumber\\
&\left. +\frac{\rho _{0}^{2}\left[ n \right] P_{\tau _p}\beta _{d, j}^{2}}{\sigma _{d}^{2}+P_{\tau _p}\beta _{d, j}} \right) \times \left( \left( \beta _{d, k}-\frac{\rho _{0}^{2}\left[ n \right] P_{\tau _p}\beta _{d, k}^{2}}{\sigma _{d}^{2}+P_{\tau _p}\beta _{d, k}} \right) \right. 
\nonumber\\
&\left. +M\left( \xi _{c, k}-\frac{\rho _{1}^{2}\left[ n \right] \xi _{c, k}\sigma _{e1, k}^{2}}{\sigma _{e1, k}^{2}+\sigma _{e2, k}^{2}+\sigma _{e3, k}^{2}+\sigma _{d}^{2}\sigma _{c}^{2}} \right) \right) ,
\\
I_{3. kn}&=A^2\beta _{r, k}\sigma _{e}^{2}tr\left( \boldsymbol{\varPhi }^H\left[ n \right] \boldsymbol{R\varPhi }\left[ n \right] \boldsymbol{R}_e \right), \label{eq33}\\
I_{4. kn}&=\sigma _{k}^{2}.
\end{align}
 }%
\begin{IEEEproof}
The proof is given in Appendix A.
\end{IEEEproof}
From the formula, we can see that with the increase of the pilot signal power \(P_{\tau _p}\) and the number of BS antennas \(N_t\), the SE will improve. When \(M=0, \) \eqref{eq30} denotes the SINR of the system without RIS assistance.

\section{Numerical Results and Discussion}
In this section, we verify the correct lines of the expressions by comparing the results of Monte Carlo simulation with theoretical calculations. At the same time, we compare and analyze the influence of \(M\), the Rician \(\kappa\)-factor and \(\rho _i\left[ n \right] \left( i\in \left\{ 0, 1 \right\} \right) \) on the SE. We denote the Rician \(\kappa\)-factor as \(\kappa =10^{1. 3-0. 003d_{br}}\), where \(d_{br}\) donates the distance between BS and RIS \cite{Shi2023Uplink}. We set the carrier frequency to 2 GHz, the channel noise per UE to \(\sigma^2=-96\) dBm, and the dimension of the RIS element to \(d_H=d_V=\lambda /2\). Moreover, the transmit pilot power is 25 dBm. Using a rectangular coordinate system, we set the positions of the BS and the RIS to (-50, 0, 30) and (0, 0, 15), respectively. The locations of the four UE is assumed to be (50, 0, 1.7), (50.5, 0, 1.7), (51, 0, 1.7) and (51.5, 0, 1.7).

Fig. 1 depicts the variations in the total SE due to different factors. The simulation results align closely with the experimental findings, indicating that the implementation of RIS significantly enhances system performance. Notably, SE tends to be stable when \(M\) tends to infinity without considering EMI. It can also be inferred from the equation that both the numerator and denominator increase simultaneously as \(M\) tends to infinity, leading the SE to approach a constant value. When considering EMI, the SE is negatively affected by the increase of \(M\). This is consistent with the result in \cite{bj2022EMImode}. Furthermore, the higher the value of \(\kappa\)-factor, the greater the SE. When the \(M\) is small, the channel aging effect is more noticeable, and when the \(M\) is large, the EMI effect is more pronounced. For example, when \(M=256\), the effect of EMI reduces SE by approximately \(8\%\), and channel aging reduces SE by approximately \(6\%\).

\begin{figure}
    \centering
            \setlength{\abovecaptionskip}{0.1cm}
\setlength{\belowcaptionskip}{0.1cm}
    \includegraphics[width=0.7\linewidth]{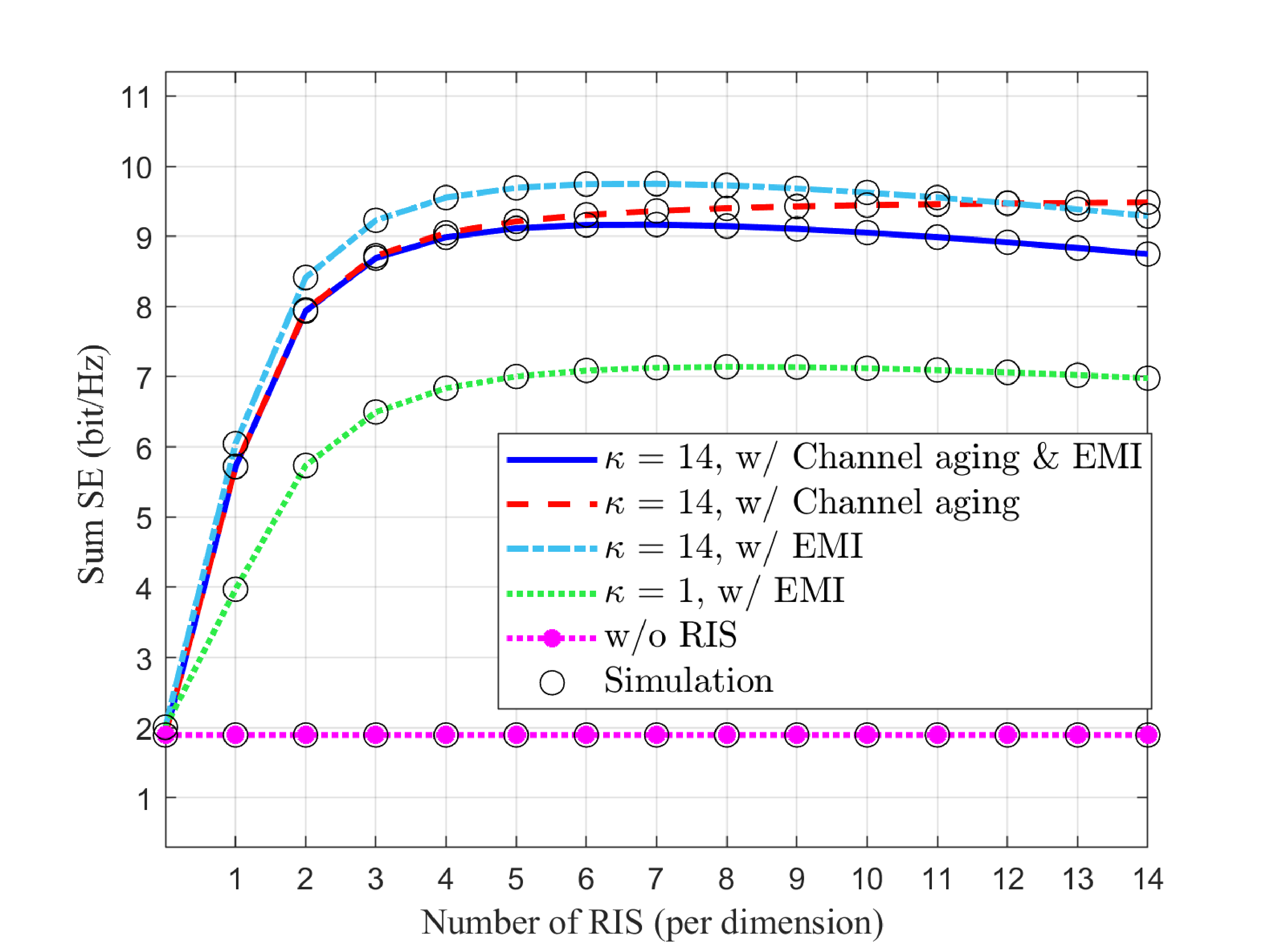}
    \caption{Sum SE`' versus $M$ and \(\kappa\) (\(N_t\) = 16, \(f_DT_s\) = 0. 001, \(\tau_c\) = 100, \(\tau_p\) = 4, \(\rho\) = 20 dB).}\vspace{-0.5cm}
    \label{fig2}
\end{figure}

\begin{figure}
    \centering
            \setlength{\abovecaptionskip}{0.1cm}
\setlength{\belowcaptionskip}{0.1cm}
    \includegraphics[width=0.7\columnwidth ]{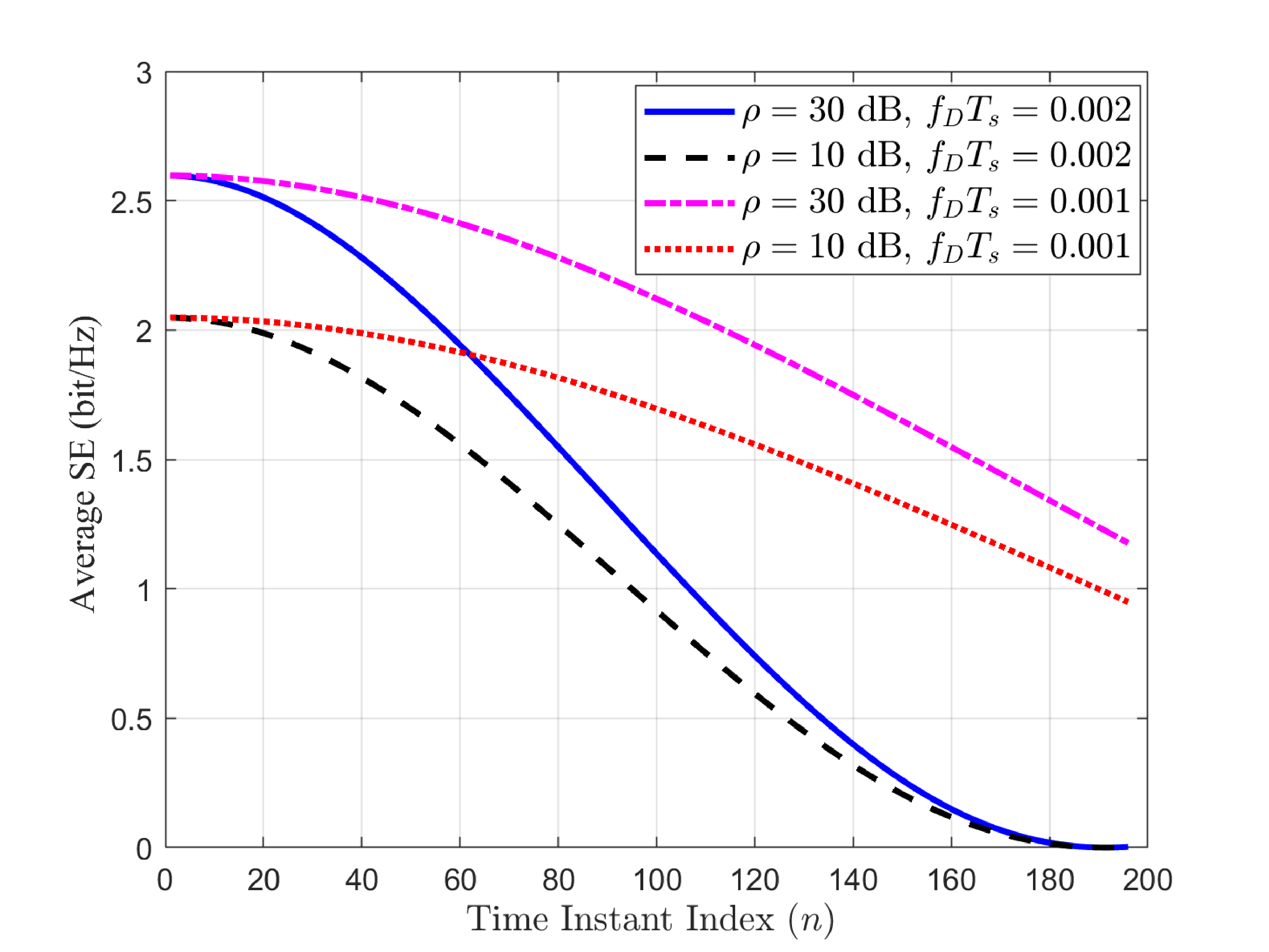}
    \caption{Average SE versus time instant (\(N_t\) = 16, \(M\) = 64, \(\tau_p\) = 4).}\vspace{-0.6cm}
    \label{fig3}
\end{figure}
We use \(f_DT_s\) to represent the impact of changes in \(\rho _i\left[ n \right]\) on SE. It can be concluded from \eqref{eq20} and \eqref{eq21} that the faster the user moves, the more serious the effect of aging on SE. Fig. 2 shows the relationship between the average SE and the transmission time under different \(f_DT_s\) and EMI. It can be seen that Increasing the value of \(f_DT_s\) will lead to a faster decline in the average SE. Similarly, an increase in EMI power will cause a decrease in SE. For \(f_DT_s=0. 002\), the total transmission time \(\tau_c\) should below 200, i. e. , \(\tau_c<200\). When the EMI power increases by 20 dB, SE decreases by \(22\%\). Additionally, we can observe that the alteration of \(\rho\) does not impact the change in time when SE decreases to 0.

\begin{figure}
    \centering
        \setlength{\abovecaptionskip}{0.1cm}
\setlength{\belowcaptionskip}{0.1cm}\includegraphics[width=0.7\linewidth]{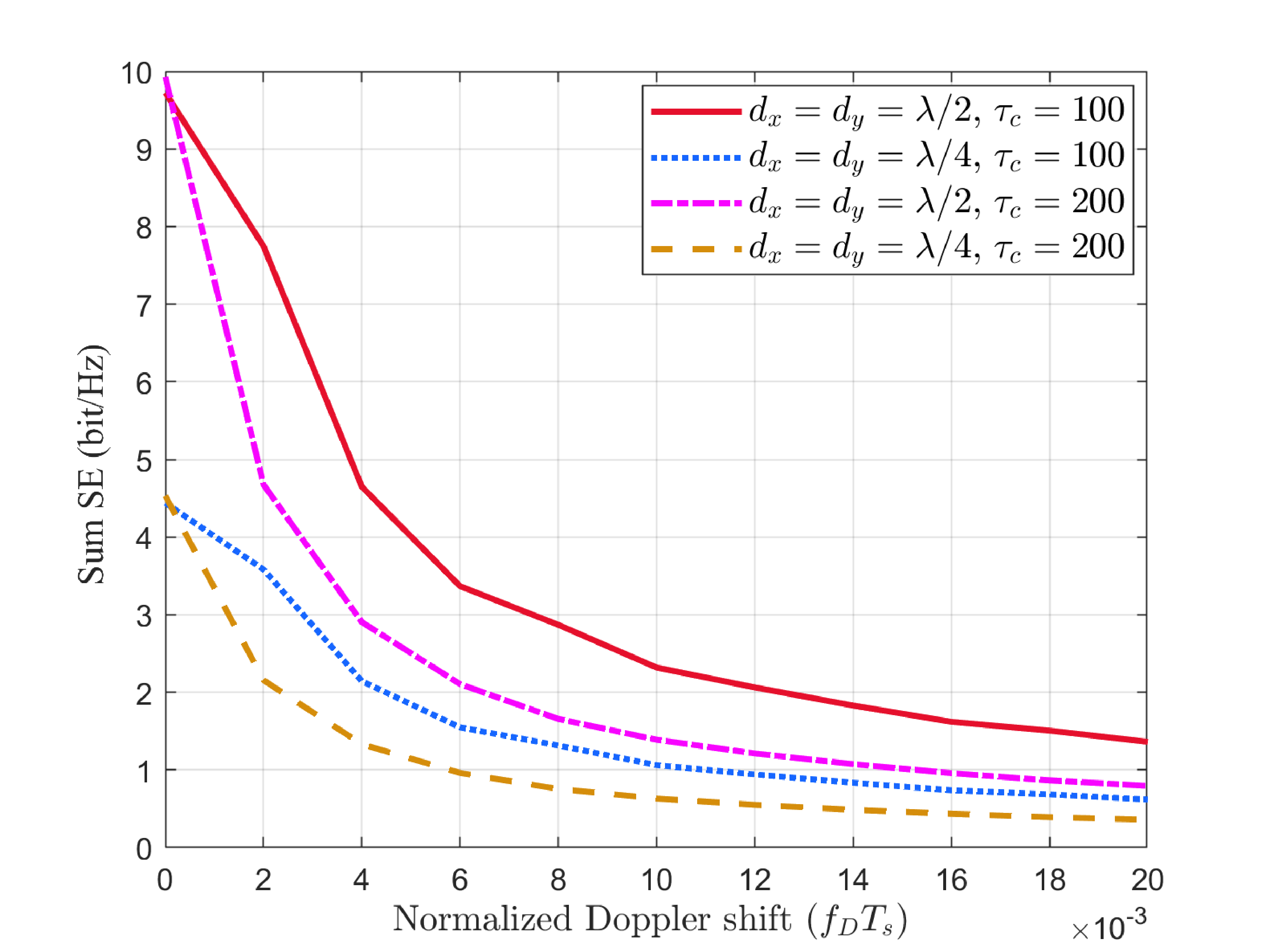}
    \caption{Sum SE versus normalized Doppler shift (\(N_t\) = 16, \(M\) = 64, \(\tau_p\) = 4, \(\rho\) = 20 dB).}\vspace{-0.6cm}
    \label{fig4l}
\end{figure}
Fig. 3 depicts the correlation of the relationship between \(f_DT_s\) and sum SE across different RIS element sizes and different transmission times. The system performance increases with the increase of RIS elements area. When \(f_DT_s=0\), \(\tau_p=4\), a longer transmission time means a larger proportion of the downlink transmission time, resulting in a higher SE of the system. However, as \(f_DT_s\) increases, longer transmission time is associated with more dramatic declines in SE due to the worsening effects of channel aging. For example, when \(f_DT_s=0. 002, M=64\) and \(\tau_c=100\), SE increases by \(116\%\) following a fourfold increase in the area of RIS elements. At this point, increasing the transmission time to \(\tau_c=200\), SE will increase by \(30\%\), indicating that expanding the area of RIS elements will restrain the negative effects of channel aging.

\section{Conclusions}\label{se:conclusion}
In this paper, we investigated the performance of the RIS-aided MISO system with the effect of channel aging and EMI. In particular, we consider the Ricican fading channel between the BS and RIS while the other paths are Rayleigh fading channels. The downlink SE closed-form expression is derived with the MRT precoding at the BS. The results indicate that the adoption of RIS can compensate for the negative effects of channel aging. However, when the number of RIS elements increases to a large number, EMI will deteriorate the system performance. In future work, we will design EMI-based precoding to mitigate the negative effects of EMI.

\section*{Appendix A}

We can calculate each part as follows.\\
(1) For \(I_{0. kn}\): According to \eqref{eq26}, because \(\boldsymbol{G}_{e, k}\left[ n \right]\) and \(\boldsymbol{f}_k\left[ n \right]\) are independent on each other, we can obtain that \(\mathbb{E}\left( \boldsymbol{G}_k\left[ n \right] \boldsymbol{f}_k\left[ n \right] \right) =\zeta \left[ n \right] \mathbb{E}\left( \boldsymbol{\bar{G}}_k\left[ n \right] \boldsymbol{\bar{G}}_{k}^{H}\left[ n \right] \right) \). 
According to the statistical properties of \(\boldsymbol{\hat{g}}_{d, k}\left[ 0 \right]\) and \(\boldsymbol{\hat{g}}_{ck}\left[ 0 \right]\), we can know that 
 {
 \setlength\abovedisplayskip{0cm}
\setlength\belowdisplayskip{0cm}
\begin{align}
&\boldsymbol{\bar{G}}_k\left[ n \right] \sim\nonumber\\
&\mathcal{N}\left( 0, \left( \frac{\rho _{0}^{2}\left[ n \right] P_{\tau _p}\beta _{d, k}^{2}}{\sigma _{d}^{2}+P_{\tau _p}\beta _{d, k}}+\frac{M\rho _{1}^{2}\left[ n \right] \xi _{c, k}\sigma _{e1, k}^{2}}{\sigma _{e1, k}^{2}+\sigma _{e2, k}^{2}+\sigma _{e3, k}^{2}+\sigma _{d}^{2}\sigma _{c}^{2}} \right) \right) ,
\end{align}
 }%
due to \( \boldsymbol{v}^H\left[ n \right] \boldsymbol{\hat{g}}_{ck}^{H}\left[ 0 \right]
=\sum\limits_{m=1}^M{v_m\left[ n \right] \boldsymbol{\hat{g}}_{ckm}^{H}\left[ 0 \right]}\). Therefore, we can obtain the expectation of \(\boldsymbol{\bar{G}}_k\left[ n \right] \boldsymbol{\bar{G}}_{k}^{H}\left[ n \right] \), because it has a Wishart distribution with complex-valued centers. Thus, we can get \(I_{0. kn}\).\\
(2) For \(I_{1. kn}\): Based on the property of the variance, we can obtain that \(Var\left( \boldsymbol{G}_k\left[ n \right] \boldsymbol{f}_k\left[ n \right] \right) 
=\mathbb{E}\left( \boldsymbol{G}_{e, k}\left[ n \right] \boldsymbol{G}_{e, k}^{H}\left[ n \right] \right) \mathbb{E}\left( \boldsymbol{f}_k\left[ n \right] \boldsymbol{f}_{k}^{H}\left[ n \right] \right) \). According to the \eqref{eq28}, we can obtain that
{
\setlength\abovedisplayskip{0cm}
\setlength\belowdisplayskip{0cm}
\begin{align}
&\mathbb{E}\left( \boldsymbol{f}_k\left[ n \right] \boldsymbol{f}_{k}^{H}\left[ n \right] \right) =\frac{\zeta ^2\left[ n \right]}{N_t}\mathbb{E}\left( \boldsymbol{\bar{G}}_{k}^{H}\left[ n \right] \boldsymbol{\bar{G}}_{k}\left[ n \right] \right).
\end{align}
 }
 \(\mathbb{E}\left( \boldsymbol{G}_{e, k}\left[ n \right] \boldsymbol{G}_{e, k}^{H}\left[ n \right] \right)\) can be written as
  {\setlength\abovedisplayskip{0cm}
\setlength\belowdisplayskip{0cm} \begin{align}
\mathbb{E}\left( \boldsymbol{\tilde{e}}_{d, k}^{H}\left[ n \right] \boldsymbol{\tilde{e}}_{d, k}\left[ n \right] \right) +\mathbb{E}\left( \boldsymbol{v}^H\left[ n \right] \boldsymbol{\tilde{e}}_{ck}^{H}\left[ n \right] \boldsymbol{\tilde{e}}_{ck}\left[ n \right] \boldsymbol{v}\left[ n \right] \right) .
\end{align}
 }%
According to the properties of the Wishart distribution, we can get \(I_{1. kn}\) due to \(\boldsymbol{v}^H\left[ n \right] \boldsymbol{\tilde{e}}_{ck}^{H}\left[ n \right] \boldsymbol{\tilde{e}}_{ck}\left[ n \right] \boldsymbol{v}\left[ n \right] =\sum\limits_{m=1}^M{\sum\limits_{l=1}^M{\exp \left( j\theta _m\left[ n \right] -j\theta _l\left[ n \right] \right) \boldsymbol{\tilde{e}}_{ck}^{H}\left[ n \right] \boldsymbol{\tilde{e}}_{ck}\left[ n \right]}}\). 
\\(3) For \(I_{2. kn}\):Because of \(\mathbb{E}\left( \left| \boldsymbol{G}_k\left[ n \right] \boldsymbol{f}_j\left[ n \right] \right|^2 \right) =\mathbb{E}\left( \boldsymbol{G}_{e, k}\left[ n \right] \boldsymbol{f}_j\left[ n \right] \boldsymbol{f}_{j}^{H}\left[ n \right] \boldsymbol{G}_{e, k}^{H}\left[ n \right] \right) \), we can obtain \(I_{2. kn}\) in the same approach as above, 
\\(4) For \(I_{3. kn}\) and \(I_{4. kn}\) :We can easily prove the \(I_{3. kn}\) and \(I_{4. kn}\).


\bibliographystyle{IEEEtran}
\bibliography{IEEEabrv,reference}

\end{document}